\documentclass[rnote]{aa} 
\usepackage{graphicx}
\usepackage{txfonts}
\usepackage{natbib}
\usepackage{booktabs}
\usepackage{tabularx} 
\bibpunct{(}{)}{;}{a}{}{,}   
\begin{document}
   \title{Unveiling the nature of IGR~J16493-4348 with IR spectroscopy\thanks{Based on observations collected at the European Southern Observatory, Chile (Programme ID  079.D-0668-B)}}


   \author{E. Nespoli
          \inst{1}
          \and
          J. Fabregat\inst{1}
          \and
          R. E. Mennickent\inst{2}
          }

   \offprints{Elisa Nespoli}

   \institute{Observatorio Astron\'omico de la Universidad de Valencia, Calle Catedr\'atico Agust\'in Escardino 7, 46980 Paterna, Valencia, Spain\\
       \email{elisa.nespoli@uv.es}
   \and Departamento de Astronom\'ia, Universidad de Concepci\'on, Casilla 160-C, Concepci\'on, Chile
     }
              


 
  \abstract
   {The International Gamma-Ray Astrophysics Laboratory (INTEGRAL) is discovering a large number of new hard X-ray sources, many of them being HMXBs. The identification and spectral characterization of their optical/infrared counterparts is a necessary step in undertaking a detailed study of these systems.}
  {In a previous paper, we presented spectral analyses and classifications of six newly discovered \emph{INTEGRAL} sources. In this paper, we extend the analysis to IGR~J16493--4348.}
   {We used the ESO/VLT ISAAC spectrograph to observe the proposed IR counterpart to the source, obtaining a $K_\mathrm{s}$ medium-resolution spectrum ($R = 500$) with a signal-to-noise ratio (S/N) $\gtrsim$ 150. We classified the source by comparing with published atlases.}
  {We spectrally classified the source as a {B0.5-1 supergiant} and estimated its interstellar extinction. We compared the extinction  derived from X-ray data with effective interstellar extinction obtained from our data, discussing the absorption component associated with the circumstellar environment. }
   {}


   \maketitle
%

\section{Introduction}


High mass X-ray binaries (HMXBs) are composed of an early-type massive star and an accreting compact object, a neutron star, or a black hole.
The majority of known systems are Be/X-ray binaries (BeXRBs), consisting of a neutron star accreting matter from the circumstellar equatorial disk of a Be star. Most of them are transient, exhibiting short and bright outbursts in the X-ray band. In the second major class of HMXBs, the supergiant X-ray binaries (SGXRBs), the compact object accretes matter from an early supergiant star through its radially outflowing stellar wind. As a consequence, most of SGXRBs are persistent systems, with $L_{X}\sim 10^{36}$ erg s$^{-1}$. \\

The \emph{INTEGRAL} survey of the Galactic plane and central regions is helping to substantially improve our knowledge of Galactic X-ray binaries \citep{bird07,bir10}. A large fraction of the newly discovered sources are heavily obscured supergiant massive X-ray binaries  \citep[first suggested by][]{rev03}, exhibiting much larger column densities ($N_\mathrm{H}\gtrsim 10^{23}$ cm$^{-2}$) than  expected along the line of sight \citep[see][]{kuulkers05}. These sources were missed by previous high-energy missions, whose onboard instruments were sensitive to a softer energy range. Optical counterparts to these obscured sources are also barely detectable because of the high interstellar extinction, $A_{V}$ being in excess of up to $\sim20$ mag.\\

In this context, infrared spectroscopy is an important tool for characterizing these systems \citep[see also][hereafter referred to as Paper I]{nes08}. With high-energy data, it helps us to identify the HMXB subclass the sources belong to and the mass-transfer process of the system, providing information about the intrinsic physics of the X-ray binary. \\

In Paper I, we presented IR spectroscopy of six \emph{INTEGRAL} sources, classifying their counterparts and estimating both their distance and interstellar extinction. In this paper, we extend the analysis including one more \emph{INTEGRAL} source, IGR~J16493-4348, located in the direction of the Norma-arm tangent region.\\

The source was discovered by \citet{gre05}. Subsequent RXTE observations by \citet{mark05} found that the mean spectrum is consistent with a heavily absorbed power law with $N_\mathrm{H} \sim 10^{23}$ cm$^{-2}$ and a photon index of 1.4. The measured flux was 1.0, 1.3, and 2.1 $\times 10^{-11}$ erg cm$^{-2}$ s$^{-1}$ in the 2--10, 10--20, and 20--40 keV energy bands, respectively.  \citet{kui05} performed Chandra imaging of the field of IGR~J16493--4348 for 4.1 ks. They detected a single point source within the $2'$ error circle of the \emph{INTEGRAL} source at R.A. =$16^h 49^m 26.92^s$, \mbox{Dec =$-43^\circ 49' 8.96''$}, with a 0.6$''$ error in each coordinate. No spectrum could be extracted from the data and \citet{kui05} noted that previous measurements of IGR~J16493--4348 by RXTE may be contaminated by another X-ray source, 1RXS~J164913.6--435527, located $\sim$6.7$'$ away.  They also pointed out that no significant source can be found at the position of the Chandra source in the optical DSS maps, indicating strong absorption in the direction of the source.
From \emph{INTEGRAL} data, \citet{hil08} discarded the previously proposed association with the free radio pulsar PSR~J1649--4349; the best-fit model of the X-ray spectrum obtained by the same authors included an absorbed cut-off powerlaw with $N_\mathrm{H} = 5.4 \times 10^{22}$ cm$^{-2}$ and $\Gamma = 0.6$. From Suzaku data, \citet{mor09} obtained $N_\mathrm{H} = 2.6 ^{+0.9}_{-0.8} \times 10^{23}$  cm$^{-2}$ and $\Gamma = 2.4$. Although having different parameters, both results are consistent with an accreting neutron star.\\

The infrared counterpart to the source was proposed by \citet{kui05}, who reported a single 2MASS source, 2MASS~J16492695--4349090, compatible with both the Chandra and Swift/XRT positions. \citet{kui05} observed the source in the $K_\mathrm{s}$ band and found a magnitude of 12, consistent with the 2MASS magnitude. No optical/IR spectra of the counterpart are available, and the {nature of the system}, although its position and X-ray behavior {suggest} that it is a HMXB system, remains unproven.\\

{In the next section, we describe the observations and data reduction. In Sect.~\ref{results}, we present the obtained spectrum, analyze its features to propose a classification, and calculate the interstellar hydrogen column density. In sect.~\ref{discussion}, we discuss our results, before concluding.} Preliminary results of our data analysis were published in \citet{nes08_b}.


\section{Observations and data analysis}    \label{observations}

Data were obtained in service mode on 2007 April 5 with the ISAAC spectrograph \citep{moo98} on UT1 at ESO/Paranal observatory. The sky was clear during the observations, the seeing was $\leq1.4''$, and the target was observed at airmass 1.09. Data were taken in the short-wavelength, low-resolution mode, in the $K$ band, with a pixel scale of 0.147$''$/pixel and a resolution of 500. Table \ref{table:logobs1} presents the observation log, including the signal-to-noise ratio (S/N) achieved {per resolution element}. Typical on-source integration times for standard stars were between 6 and 10 seconds.\\

\begin{table}[]
      \caption{VLT/ISAAC observation log. } 
              \begin{center}
         \label{table:logobs1}
            \begin{tabular}{lc}
                       \hline
             \hline
             \noalign{\smallskip}
Start time  (UT)            & 2007-04-05  09:38\\
Exp. time (s)  & 360\\
S/N  & 150\\
R ($\lambda/\Delta\lambda$)  & 500\\
IR counterpart  & 2MASS J16492695-4349090\\
$K$ mag  & 11.9\\
Reference  & Kuiper et al. (2005)\\
            \noalign{\smallskip}
            \hline
\end{tabular}
       \end{center}
\label{lines}
\end{table}

Data reduction was performed using the IRAF\footnote{IRAF is distributed by the National Optical Astronomy Observatories which is operated by the Association of Universities for Research in Astronomy, Inc. under contract with the National Science Foundation} package, following the standard procedures for IR spectra, described in detail in Paper I. Removal of atmospheric features was accomplished by employing a G2 V and a A3 V spectrum, following a strategy similar to that outlined by \citet{clark2000}. For a detailed description of the correction procedure, we refer to Paper I. Between the standard telluric spectra and the scientific one, efforts were made to ensure differences of no more $\sim$0.01 in the airmass.


\section{Results}    \label{results}

We present our spectral classification and analysis results for IGR~J16493-434. Our data is the first available spectrum of this source. The analysis will be qualitative, based on comparison with available NIR spectral atlases \citep{hanson96,hanson05}. \\

\begin{figure*}[ht!]
 \centering
\includegraphics[width=18cm]{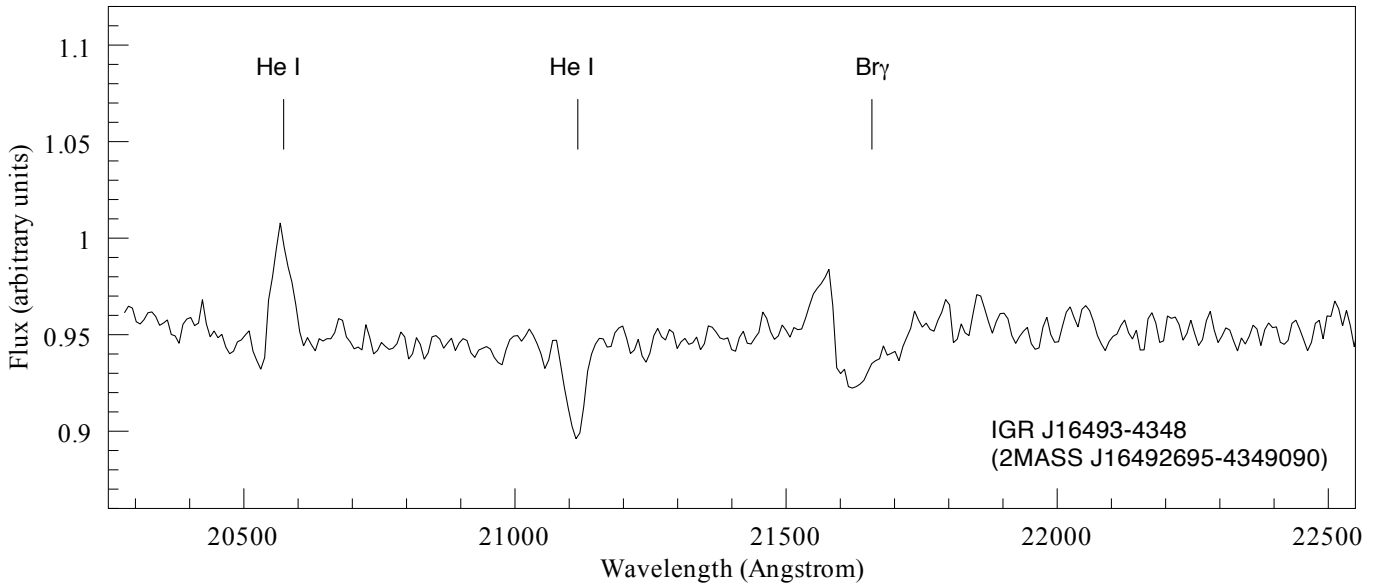}
   \caption{\small{$K_\mathrm{s}$ spectrum for 2MASS~J16492695--4349090, the infrared counterpart of IGR J16493--4348. The positions of identified spectral features are indicated by solid lines.}}
            \label{fig:spec}
   \end{figure*}

The {processed} $K$-band spectrum is shown in Fig.~\ref{fig:spec}, while its most significant features are reported in Table~\ref{lines}.

\begin{table}[!ht]
\caption{$K$-band line identifications.}
\begin{center}
\resizebox{8.7cm}{!}{
\begin{tabular}{lccr}
\hline
\hline
  \noalign{\smallskip}
 Feature           & Transition & Wavel. [\AA]  & EW [\AA]      \\ 
   \noalign{\smallskip}
\hline 
  \noalign{\smallskip}
\ion{He}{I}       & (2s$^1$S - 2$p^1$P$^o$), 20\ 581 \AA             & 20\,571  &     -2.7$\pm$0.4    \\ 
\ion{He}{I}       & (3p$^3$P$^0$ - 4s$^3$S), 21\,120 \AA             &                   &    \\ 
                          & + (3p$^1$P$^0$ - 4s$^1$S), 21\,130 \AA & 21\,112   & 2.0$\pm$0.4\\
Br$ \gamma$ & (4$^{2}$F$^{o}$  - $7^{2}G$), 21\,661            & 21\,655  &      2.7$\pm$0.2   \\
  \noalign{\smallskip}
  \hline
\end{tabular}}
\end{center}
\label{lines}
\end{table}

Errors in the equivalent widths were calculated as deviations from the mean values, measured using  different estimates of the continuum level. Some of the identified features exhibit a 5--10 \AA\ displacement with respect to the nominal values, which is consistent with the instrumental resolution. \\

Although it has been pointed out \citep{hanson96} that equivalent widths may vary between stars of the same spectral type, we report our measurements for completeness. 
\subsection{Spectral analysis and classification}

The spectrum exhibits a strong emission \ion{He}{I} 20\,581 \AA~ line, \ion{He}{I} 21\,126 \AA~in absorption, and a relatively strong absorption Br$\gamma$ 21\,661 \AA~line. {We also tentatively detected Br$\gamma$ in faint emission, which might be a signature of an accretion disk around the compact object. However, the detection is very weak, and we cannot exclude it being a residual artifact from the telluric-line removal procedure, which in this region is very critical.} The atomic transitions detected are the typical ones of OB stellar  spectra. In particular, \ion{He}{I} 20\,581 \AA~is a prominent feature for supergiant stars and is observed in emission in B type supergiants. The He I 21\,126 \AA\ line is present in late O - early B spectra.  By comparing the relative strength of the identified lines with those from the atlas of \citet{hanson96}, we estimated the spectral type to be {B0.5-1 Ia-Ib}. With information provided by X-ray data \citep{hil08}, our results thus permit us to classify the system as a neutron-star SGXRB.

\subsection{Reddening and distance estimation}

As in Paper I, we proceeded to calculate the extinction and the distance to characterize the system.\\

From the proposed spectral classification, we obtained the intrinsic colors $(J-K)_{0}$ from \citet{weg94}. Apparent colors were calculated from 2MASS photometry, properly transformed by means of the formula from \citet{car01}\footnote{In its updated version at http://www.astro.caltech.edu/$\sim$jmc/2mass/\\ v3/transformations/} to the \citet{bebe88} homogenized photometric system to estimate the infrared color excess $E(J-K)$. We assumed the mean extinction law ($R_{V}= 3.1$), from which $A_{V}/E(J-K) = 5.82 \pm 0.10$ \citep{rieke85}, obtaining the total measured visual extinction $A_{V}$, and the corresponding hydrogen column density value from $N_\mathrm{H}/A_{V} = 1.79 \pm 0.03 \times 10^{21}$ atoms cm$^{-2}$ mag \citep{pred95}. We were thus able to compare the measured interstellar value of $N_\mathrm{H}$ with that inferred from X-ray data. In our calculation, we estimated errors by means of error propagation. The error in the final value of $N_\mathrm{H}$ is mainly caused by errors in the infrared colors and the transformation between the two photometric systems. {In Table~\ref{tab:reddening}, we present the results of our calculations, namely the spectral classification of the IR source, the intrinsic infrared colors, 2MASS photometry, the infrared excess, the hydrogen column density obtained from X-ray published measurements (most recent results), and the effective interstellar column density obtained from our work.}
\\

{We also estimated the distance to the source, by means of the relation $M_{K} = K + 5 - 5$ log $d - A_{K}$, employing intrinsic colors and 2MASS magnitudes (see Paper I for procedures and references). For this system, the huge uncertainty in the intrinsic colors directly affects our estimate of distance, which may have any values in the range 6--26 kpc. This range is too broad to allow us to determine the precise location on the source in the Galaxy, and in particular to confirm its association with any Galactic spiral structure. Our distance measurement is thus not shown in Table \ref{tab:reddening}.}

\begin{table}[!ht]
\caption{{Results from reddening estimation.}}
\begin{center}
\begin{tabular}{lr}
\hline
\hline
  \noalign{\smallskip}
Spectral type & B0.5-1 Ia-Ib\\
 $(J-K)_{0}$ (mag) & -0.12\\
 $(J-K)_{2MASS}$ (mag) & 2.66\\
 $E(J-K)$ (mag) &  2.81\\
 $N_\mathrm{H}$ from X-ray data (10$^{22}$ cm$^{-2}$) & $26^{+9.4}_{-7.9}$\\
Interstellar $N_\mathrm{H}$ (10$^{22}$ cm$^{-2}$) & $2.92 \pm 1.96$\\
  \noalign{\smallskip}
  \hline
\end{tabular}
\end{center}
\label{tab:reddening}
\end{table}

\section{Discussion and conclusions}     \label{discussion}

Using $K$-band spectroscopy, we classified the counterpart to IGR J16493--4348  by comparison with published atlases. We found that the system  has a supergiant companion, estimating its type to be {B0.5--1 Ia-Ib}. Combined with information from X-ray data, our results permit us to classify the system as a SGXRB.\\

This work allowed us to calculate the extinction from IR data. Previous works have pointed out \citep[see][]{kuulkers05,cha07} that \emph{INTEGRAL} {is detecting} a new class of highly obscured supergiant HMXBs. The origin and position (around the compact object only, or enveloping the entire system) of the absorbing material remain unclear, and only multiwavelength studies can help us to address the problem, by distinguishing between the absorption in X-ray and in the IR/optical bands. \\

 We calculated the effective interstellar extinction $A_{V}$ and converted it into a hydrogen column density of $N_\mathrm{H} = 2.92\pm1.96 \times10^{22}$. This value is compatible with estimates of the weighted average neutral hydrogen density in a cone of radius $1^{\circ}$ along the line of sight of IGR J16493--4348 \citep[$\sim$ 1.4--1.8$\times10^{22}$ cm$^{-2}$,][]{kal05,dic90}. Our results were compared with the values obtained from X-ray data. If the two derived values were compatible within their corresponding errors, we would have to consider two possible scenarios: either the source of absorption is just the interstellar medium, or there is a contribution from an extensive envelope around the whole binary system, if the extinction is systematically higher by some orders of magnitude than the estimated interstellar value. In contrast, if the reddening (measured from the IR colors) were low compared to the measured $N_\mathrm{H}$ from X-rays, this would imply that there is an additional source of extinction, which only affects the compact object in which the X-ray emission originates and can be assumed to be the absorbing material around it.\\
 
In our case, the extinction measured at high energy was found to be one order of magnitude higher than that obtained from IR data, which indicates that the material absorbing in the X-ray is concentrated around the neutron star. The same results were found in Paper I for the following systems: IGR J16465--4507, IGR J16479--4514, AX J1841.0--0536, and IGR J19140+0951.\\

The so-called highly absorbed IGR sources are usually identified to be those for which the measured $N_\mathrm{H}$ is $\gtrsim$$10^{23}$ cm$^{-2}$, \emph{i.e.}, one to two orders of magnitude higher than the assumed Galactic value of \mbox{$\sim10^{22}$ cm$^{-2}$} \citep{kuulkers05}. As in Paper I, the comparison of X-ray data, which are sensitive to the absorption caused by the environment of the compact object, with infrared data, which indicate in general only the radiation absorbed by the interstellar medium, was a powerful, alternative criterion for identifying this class of highly absorbed sources.  \\

\section{Conclusions}

Using $K$-band spectroscopy, we have extended the analysis presented in Paper I to another \emph{INTEGRAL} source, IGR~J16493-4348, which is part of {the} new class of highly-obscured SGXRBs. We have presented the first available IR spectrum of the counterpart and found that:

\begin{itemize}
\item[-] the proposed optical counterpart was confirmed and classified as {B0.5-1 Ia-Ib};
\item[-] the system was thus classified as a SGXRB;
\item[-] the comparison between $N_\mathrm{H}$ obtained from X-ray data and the interstellar extinction from our data showed the presence of an absorbing envelope, tightly confined to the vicinity of the compact object.
\end{itemize}

  \begin{acknowledgements}
The work of EN and JF is supported by the Spanish Ministerio de Educaci\'on y Ciencia, and FEDER, under contract AYA 2007-62487. This work has been partly supported by the Generalitat Valenciana project of excellence PROMETEO/2009/064. EN acknowledges a ``V Segles'' research grant from the University of Valencia. REM acknowledges support by Fondecyt 1070705, the Chilean Center for Astrophysics FONDAP 15010003 and the BASAL Centro de Astrof\'isica y Tecnolog\'ias Afines (CATA) PFB--06/2007. 
 
      \end{acknowledgements}

\bibliographystyle{aa}

\end{document}